\documentstyle[epsfig,12pt]{dubna98}
\def\Tr{\mathop{\mbox{Tr}}\,}
\def\fnote#1#2{\begingroup\def\thefootnote{#1}\footnote{#2}\endgroup}

\begin{document}
\begin{center}

\begin{flushright}
BUTP-98/17\\
Bern, July 1998
\end{flushright}

\vspace{0.8cm}
{\bf DISCUSSION ABOUT THE ORTHOPOSITRONIUM DECAY RATE AND ANALYSIS OF SOME 
$O(\alpha^2)$ CONTRIBUTIONS\\}

\vspace*{1cm}

{V. ANTONELLI~\fnote{\dag} {Extended version of the talk given at the 
International Workshop 
"HADRONIC ATOMS AND POSITRONIUM IN THE STANDARD MODEL" \, , DUBNA, 
26-31 MAY 1998}}\\
{\it Institute for Theoretical Physics ~-~ University of Bern,\\}
{\it Sidlerstrasse 5 \, CH-3012 Bern \, Switzerland\\}
{e-mail address 
antonell@itp.unibe.ch}\\

\end{center}

\vspace*{.5cm}

\begin{abstract}{\small  I consider the problem of the evaluation of the 
orthopositronium decay rate, including second order radiative corrections.
I present a brief theoretical discussion of this problem and a review of the 
results available in literature, and, then, I analyze some $O(\alpha^2)$ annihilation-type contributions, recently computed.}

\end{abstract}

\vspace*{1cm}

\section{Introduction}

Positronium is a bound state of an electron and a positron. It is 
essentially a pure quantum electrodynamical system; in fact the effect of strong and weak interactions can be safely neglected at the level of accuracy
at which we are interested.

Its study has always been considered an interesting test of the
accuracy of QED calculations and especially of the formalism used to describe
bound states in quantum field theory.

The lifetime of this atom can vary in a quite large range from $10^{-10}$ to 
$10^{-7}$ s, according to its spin state. 
In fact it can exist in two different spin states: a singlet $S=0$ state, 
called parapositronium, and a triplet $S=1$ state, called 
orthopositronium.

Parapositronium decays mainly into a couple of photons. The triplet state, 
instead, decays electromagnetically into an odd number of photons greater than
one, because decays into an even number of photons are forbidden by charge
conjugation invariance and one photon decay would violate the energy momentum 
conservation. Hence, the channel with three photons in the final state is the 
dominant decay mode for the orthopositronium and one can restrict the analysis
to this channel at the 10 ppm level of accuracy.

We will focus our attention to the analysis of orthopositronium decay width.

Despite of the fact that different precision measurements and theoretical 
computations of this quantity have been performed in the past years, there is 
still a certain indetermination in the results and one cannot exclude a 
possible discrepancy between theory and experiment. This discrepancy could
eventually indicate also a problem in the formalism used to study this
electromagnetic bound state system.

The more precise experimental determinations of orthopositronium decay rate,
obtained by the Michigan University group \cite{mic1,mic2}, are the following:

\begin{equation}
\lambda^{exp}_{Ops} = 7.0514 \pm 0.0014 ~ \mu s^{-1} ~ ; \:  
\lambda^{exp}_{Ops} = 7.0482 \pm 0.0016 ~ \mu s^{-1} ~ .
\end{equation}   

The first measurement was performed by stopping in a gas the positrons
produced by the $\beta$ decay of a radioactive source, in such a way to
produce positronium, and looking at the annihilation $\gamma$ rays produced 
by the O-ps decay. The spectrum obtained was then fitted to an exponential 
function, taking care to begin the fit at a late enough time, so that the 
perturbed singlet component was not present anymore. 
Finally, the values of $\lambda$ 
obtained at different gas densities were extrapolated linearly to zero 
density. In the second Michigan experiment \cite{mic2}, instead, a different 
technique was used and orthopositronium was produced in evacuated MgO-lined 
cavities. 

A more recent Japanese experiment \cite{giap} found a quite 
different value 
\begin{equation}
\lambda^{exp}_{O_{ps}} = 7.0398 \pm 0.0025 \pm 0.0015 ~ \mu s^{-1} ~,   
\end{equation}   
where the first error is statistical and the second systematic.

There had been also other measurements of the decay width (we can remember in 
particular the value  $\lambda^{exp}_{O_{ps}} = 7.031 \pm 0.007 ~ \mu s^{-1}$ 
obtained by the Mainz experiment \cite{mainz} using an evacuated aluminum
cavity), but they were all characterized by higher values of the error.
\vspace*{.5cm}

The theoretical expression for the orthopositronium decay rate, in the 
approximation of considering only the decay into three photons, can be 
written in the form ~\fnote{\dag}{Here we are adopting the 
convention to collect an explicit power of $(\frac{\alpha}{\pi})^n$ in front of
the contribution of order n to the decay width, to be consistent with what is
usually done in literature. We would like, however, to stress that this
convention, justified by the fact that this is the usual natural scale of the
relativistic corrections, can be misleading. In fact, in this 
way, the coefficients of the higher order radiative corrections appear
unnaturally big.}
\begin{eqnarray}
\lambda^{th}_{O_{ps}} &=& \lambda_0 \, \left[\,1\,+ A\, \frac{\alpha}{\pi}\,
- \frac{1}{3}\, \alpha^2 \ln \frac{1}{\alpha}  \,+ B\,
\left(\frac{\alpha}{\pi}\right)^2\,- \frac{3}{2}\, \frac{\alpha^3}{\pi} \ln^2
\frac{1}{\alpha}\,+O((\alpha /\pi)^3)\,\right]\,  ,
\label{masterformula}
\end{eqnarray}
where the lowest order decay rate $\lambda_0$, first obtained by Ore and
Powell \cite{Ozero}, is given by 
\begin{equation} 
\lambda_0\,=\, \alpha^6 ~ \frac{m c^2}{\hbar} ~ \frac{2 \, \left(\pi^2 -9
  \right)}{9 \pi}\, =\, 7.211169 \pm 0.000004 \mu s^{-1}\, .
\label{mancava}
\end{equation}

The coefficient A has been computed by many authors \cite{Oalfa,Oalfab,OalfaAd,Adsolo}
and its most accurate determination $A \,=\, -10.286606 \pm 0.000010$ has been
found in \cite{Adsolo}. The constants $\frac{-1}{3}$ and $\frac{-3}{2 \pi}$
of the logarithmic terms have been found respectively in \cite{Oalfab,
  OalfaAd, log1} and \cite{log2}. Using these known results, one gets:
\begin{equation}\ 
\lambda_{Ops}^{th} \, = \, 7.038204 \pm 0.000010 \mu s^{-1} \,+ \lambda_0 \,
\left[B \left( \frac{\alpha}{\pi}\right)^2 \, +
  O\left(\frac{\alpha}{\pi}\right)^3\right] \, . 
\label{fulla2}
\end{equation}

We can see from the last expression that there could be a discrepancy between 
the theoretical and the experimental results. In fact, putting $B=0$ in 
eq.~(\ref{fulla2}), we obtain a theoretical value at $O(\alpha)$, with the
inclusion of logarithmic terms up to $O(\alpha^3 \ln^2 \alpha)$, which differs
by the experimental results of the Michigan group \cite{mic2,mic1} by
$6.2 \, \sigma$ and $9.4 \,  \sigma$. To reproduce these results one should 
have a value of the coefficient $B \simeq 250$, that would be unnaturally
big. On the other hand, $B \simeq 40$ would be enough to reproduce the result
of the Japanese experiment \cite{giap}. Hence, up to now, one cannot
discriminate whether the so called ``orthopositronium decay width puzzle'' is a
theoretical or an experimental problem. It is clear, in any case, that a
complete $O(\alpha^2)$ calculation is needed.  

Some second order contributions to the decay rate have been already computed 
by many authors, but some other still need to be evaluated. 

Here we focus our attention on some second order annihilation type radiative 
corrections that have recently been computed in \cite{io}.

\section{Analysis of the bound states in quantum field theory}

The study of a bound state like positronium is a complex problem that can be
faced only using difficult techniques of quantum field theory and making some
appropriate approximations \cite{bodyenrev}. 

The first simplification of the problem can be obtained considering
positronium as a two body system. The development of a consistent relativistic
two body formalism for bound state calculations is mainly due to the works of
Schwinger \cite{Schwinger} and Bethe and Salpeter \cite{Betesalp}.

To study the bound states of a two body system, we have, essentially, to look
for the poles of  a four point Green function, that, for fermions, is defined 
in the usual quantum field theory way as ~\fnote{\dag}{Here we are using, for
  simplicity reasons, a two particle formalism, instead of a
  particle-antiparticle one.
In the specific case of positronium, one should of course charge conjugate one
of the particles of our system.}:
\begin{equation}
G \left(x_1, x_2,  x_3, x_4 \right) \,=\, \langle 0 | T \left(\psi (x_1) \psi
(x_2)  \bar{\psi} (x_3) \bar{\psi} (x_4) \right)\, |0 \rangle \, .
\end{equation}

This Green function must obey the Bethe-Salpeter equation
\begin{eqnarray}
G \left(x_1, x_2,  x_3, x_4 \right) ~ = ~ S_F^{(1)} \left(x_1, x_3\right) \, 
S_F^{(2)} \left(x_2, x_4\right)   &+& \nonumber\\ 
+\, \int dx_5 ~ dx_6 ~ dx_7 ~ dx_8  ~S_F^{(1)} \left(x_1, x_5\right) ~ 
S_F^{(2)} \left(x_2,  x_6\right) &\times&K \left(x_5,  x_6, x_7, x_8\right)\, G
\left(x_7,  x_8, x_3, x_4\right),
\label{betesalp}
\end{eqnarray}
~\fnote{\dag \dag}{Notice that, to simplify the notation, here and in the
  following we are denoting by dx the four dimensional integral $d^4 x$} 
where $S_F$ is the Feynman propagator.

This equation can also be represented graphically, like in fig. 1 .

In eq.~(\ref{betesalp}) we have denoted by K the sum of all the two particle
irreducible graphs, that is usually called the Bethe-Salpeter kernel.

One can use the translational invariance properties of the system and express
the Fourier transformed Green function in terms of only three variables:
\begin{eqnarray} 
\int dx_1~ dx_2~ dx_3~ dx_4~ exp\left[ i \left(p_1 x_1 +p_2 x_2 - p_3 x_3- p_4
 x_4\right) \right]&\times& G\left(x_1, x_2, x_3, x_4\right) =\nonumber\\
 \,=\, (2 \pi)^4 \delta^{(4)} \left(p_1+p_2-p_3-p_4\right) ~ G (p,q,P)\,& , & 
\end{eqnarray}
where 
\begin{equation} 
p \,=\, \frac{1}{2} \, (p_1-p_2) \, ; \, q \,=\, \frac{1}{2} \, (p_3-p_4) \, ;
\, P \,=\, p_1+p_2 \,=\, p_3+p_4 \, . 
\label{def4point}
\end{equation}

The next step is the introduction of the two-particle (two-antiparticle) 
physical states $|i, P \rangle^{\pm}$ of mass $M_i$ and four momentum $P_i~=~
(E_i,\vec{P})$, $E_i ~=~\sqrt{\vec{P}^2+M_i^2}$.

Using these states, we can define the Bethe-Salpeter wave functions in the
following way: 
\begin{eqnarray}    
\chi_{i,P}^{\pm}(p)&=&\int d (x_1-x_2) ~e^{i p \cdot (x_1-x_2)}~ \langle
0~|T \left(\psi^{(1)}(x_1)~\psi^{(2)}(x_2)\right) |i,P \rangle^{\pm}~ e^{i P_i
\cdot \frac{1}{2} (x_1+x_2)}
\nonumber\\
\bar{\chi}_{i,P}^{\pm}(p)&=&\int d (x_3-x_4) ~e^{-i q \cdot (x_3-x_4)}
~^{\pm}\langle i, P|T \left(\bar{\psi}^{(1)}(x_3)~\bar{\psi}^{(2)}(x_4)\right)
 |0 \rangle~ e^{-i P_i \,
\frac{1}{2} (x_3+x_4)} ~ .
\label{BSWfunct}
\end{eqnarray}

One can show that the Green function G(p,q,P) can be written as a sum 
containing the Bethe-Salpeter wave functions $\chi^{\pm}$ and  
$\bar{\chi}^{\pm}$ and this sum has some discrete singularities for specific 
values of $P$ (for which $P_0=\pm \sqrt{\vec{P}^2+M_i^2}$ ).

Using Bethe-Salpeter equation, we get an equation for the residua 
of the poles of the Green function, that gives:
\begin{equation}
\chi^{\pm}_{i,P} (p)~=~S_F \left(\frac{1}{2} P + p\right)~S_F
\left(\frac{1}{2} P - p\right)~\int \frac{d^4 s}{\left(2 \pi\right)^4}~K
(p,s,P) \chi_{i,P}^{\pm} (s) \, .
\label{reschi}
\end{equation}

The values of $P$ giving non trivial solutions of eq.~(\ref{reschi}) are the 
energy-momenta of the bound states coupled to the two particle-antiparticle 
states.

One usually looks for perturbative solutions of this equation, by writing the 
kernel $K$ as the sum of a piece $K_0$, whose solution is known,  and a 
remainder $\Delta K$.

We can also take advantage of the fact that a QED bound state, like
 positronium, can be formed only if the relative momentum of the $e^+e^-$ pair
is small, since the binding force is weak. The expectation value of the 
relative momentum in the bound state is of the order $p \simeq \alpha ~ m$, 
where $m$ is the mass of the electron. Hence we are allowed to make a non 
relativistic approximation and to distinguish a ``large component'' from a 
small one in the solution $\psi$ of Dirac equation:
\begin{equation} 
\Psi ~=~ \left(\begin{array}{c}\chi\\ \\\frac{\vec{p} \cdot
      \vec{\sigma}}{p_0+m}~ \chi\end{array}\right)\, .
\label{positwf}
\end{equation}

The first order Bethe-Salpeter kernel contains in our case the product of a 
photon propagator $D_{\mu \nu}$ times two Dirac gamma matrices $\gamma_{\mu}$ 
and $\gamma_{\nu}$ and the only term of this product connecting the
large-large components is the term $D_{00} \gamma_0 \gamma_0$. 
Hence, working in the Coulomb gauge, we can choose the following expression 
for the unperturbed kernel of the Bethe-Salpeter equation:
\begin{equation}
K_{0} (\vec{p}^{~'},\vec{p})~=~ \frac{4 \pi \alpha
  i}{(\vec{p}^{~'}-\vec{p})^2}~\gamma_0 \times \gamma_0 \, . 
\label{Coulkern}
\end{equation}

In this way, we get for the large-large components the well known
Schr\"odinger equation with a coulombic potential. 
The lowest order expression of the ground state wavefunction for the 
orthopositronium can, therefore, be written as 
\begin{equation}
\Psi(p) ~=~ 2 \pi ~ \sqrt{2 m} ~\delta(p_0)~\phi(\vec p) ~
\left[ \begin{array}{ccc} 0 & \vec{\sigma} \cdot \vec{\epsilon}_m \\
                          0 & 0 \\
\end{array}
\right]  \, 
\label{groundortopos}
\end{equation} 
where 
\begin{equation}
\phi(\vec p)~=~ \phi_0 \frac{8 \pi \gamma}{(\vec{p}^{~2}+\gamma^2)^2}~ ; ~ 
\phi_0~=~\sqrt{\frac{\gamma^3}{\pi}} ~;~ \gamma~=~\frac{m \alpha}{2}\, . 
\end{equation}

This is the expression that we will use to compute the orthopositronium decay 
rate.

\section{The orthopositronium decay width}

At lowest order, the orthopositronium decay width can be obtained simply
 considering the diagram of fig. 2. 

We can easily see that it is given by
\begin{equation}
\Gamma = \frac{1}{2 M} \int \frac{d^3 k_1}{2 \, \omega_1 \, \left(2
    \pi\right)^3} 
\frac{d^3 k_2}{2 \, \omega_2 \, \left(2 \pi\right)^3} \frac{d^3 k_3}{2\,
    \omega_3 \, \left(2 \pi\right)^3} ~ \left(2 \pi\right)^4 \delta^{(4)}
    (P-k_1-k_2-k_3) \sum_{\varepsilon_1, \varepsilon_2, \varepsilon_3}~\frac{1}{3}~
\sum_{\varepsilon_m} \frac{1}{3!} \left|{\cal M}_0\right|^2 \, ,
\label{order0dec}
\end{equation}
where $M$ is the orthopositronium mass, $P \equiv 
(M,0,0,0)\, \simeq (2 m, 0,0,0)$  the rest-frame orthopositronium
energy-momentum vector and $\omega_i \equiv k_i^0 = \left|\vec{k}_i\right|$
the energies of the final state photons.
In eq.~(\ref{order0dec}) we have denoted by $\varepsilon_i$ and 
$\varepsilon_m$, respectively, the polarization states of the final photons 
and of positronium and with ${\cal M}_0$ the matrix element for the 
graph of fig. 2.

One recovers, in this way, the well known lowest order value for the decay
width, already found in \cite{Ozero} (see eq.~(\ref{mancava})).

At the next order of perturbation theory, we must take into account different
classes of diagrams, like self-energy and vertex corrections and the 
radiative corrections given by the two graphs of fig. 3, usually denoted as 
annihilation and binding diagrams.   

The last one is particularly important for many reasons. In fact one can
easily show that the matrix element ${\cal M}_B$ for this graph can be written
in the form 
\begin{equation}
{\cal M}_B = {\cal M}_0 \left(1-\ 3\ \frac{\alpha}{\pi}\right)\, ,
\label{oalfabind}
\end{equation}
where ${\cal M}_0$ is the matrix element of the lowest order graph of  fig. 2. 

Hence, it seems that this diagram gives contribution to the decay width, not 
only at
order $\alpha$, but also at order zero. This unusual fact can be explained
considering that, when we add an additional binding photon to the original
lowest order diagram of fig. 2, we are considering also the
possibility that this binding photon is a coulombic one. On the other hand, in
the determination of the positronium wave function one has already taken into
account the exchange of any number of coulombic photons between the electron
and the positron lines. It is, therefore, clear that one must subtract from the
contribution of the last graph of fig. 3 the part corresponding to the
exchange of a Coulomb photon and this correspond to subtract the term of order
zero from the expression of eq.~(\ref{oalfabind}). In this way, one gets the
following result for the matrix element of 
the ``subtracted binding diagram" $\frac{\alpha}{\pi} {\cal M}_{B}^{'}$ 
\footnote{Note that here and in the rest of the paper we write 
explicitly in the formulas the powers of $\alpha /\pi$ appearing in all the 
amplitudes; on the contrary, we omit them in the text, with the exception of 
this line. Note also that we will not write any power of $\alpha /\pi$ for the two unsubtracted amplitudes $M_{B}$ and $M_{AB}$, since they contain terms of different order in $\alpha /\pi$.}:

\begin{equation}
\frac{\alpha}{\pi} {\cal M}_B^{'} = - 3\ \frac{\alpha}{\pi} \ {\cal M}_0 \, . 
\label{suboalbind}
\end{equation}

The binding diagram is particularly important also because it gives a
contribution to the decay rate, which is bigger than the $90 \%$ of 
the total $O (\alpha)$ radiative corrections \cite{OalfaAd}.
The annihilation diagram gives a contribution \cite{Oalfaann} that has the
same  sign of the binding contribution, but is, anyway, smaller than  $10\%$ 
of the total $O(\alpha)$ correction. The sum of all the vertex corrections 
has the same order of magnitude and the opposite sign, and the self energy 
correction is even smaller \cite{OalfaAd,selfenvert}.

As already said, we must consider also the second order radiative
corrections, some of which have been already computed. 

A first relevant contribution of this order, first evaluated in
\cite{sumsquares} and recently updated in \cite{Adsolo}, is given by the sum 
of the squares of all the first order amplitudes. 
They give a contribution equal to $28.860 \pm 0.002$ to the coefficient $B$ of
eq.~(\ref{masterformula}). 

A second contribution \cite{radscattbloc} to $B$, equal to $9.0074 \pm
0.0009$, comes from the radiative corrections to the light-light scattering
block. The inclusion of the vacuum polarization corrections to the first order
graphs enhances the value of the coefficient $B$ of  $0.964960 \pm 0.000004$ as
proved in \cite{vacpolar}. 

Considering the decay channel into five photons, one gets \cite{5fot} an
additional contribution to $B$ equal to $0.187 \pm 0.011$.  

The very important second order relativistic corrections have been studied by
different authors. Khriplovich and Milstein \cite{2ordrel} have found a big
contribution to the coefficient $B$, equal to $46 \pm 3$, in agreement with
the result of Faustov et al.\cite{Faustovetal}. 
Quite a different result have been found with a different approach, in the 
second paper of \cite{log2}, by Labelle et al., that, using the so called 
``Non Relativistic Quantum Electrodynamics'' have got for this contribution 
the value $24.6$. 

Generally speaking, we can write the matrix element for the sum of all the
diagrams contributing up to $O(\alpha^2)$ in the following way:
\begin{equation}
{\cal M} = {\cal M}_0+\, \frac{\alpha}{\pi} \left({\cal M}_{B}^{'}+\ {\cal
    M}_{A}+\, {\cal M}_1\right)+\, \left(\frac{\alpha}{\pi}\right)^2 \, 
\left({\cal M}_{AB}^{'}+\, {\cal M}_{AR}+\, {\cal M}_2 \right)+\,
    O(\alpha^3)\, ,
\label{complete2order}
\end{equation}
  where $M_1$ represents the sum of all the first order amplitudes with the 
exceptions of the first order annihilation diagram     , 
denoted by $M_{A}$, and  the subtracted binding amplitude, $M_{B}^{'}$. 
The second order annihilation type corrections  are given by the 
subtracted binding diagram, $M_{AB}^{'}$ (fig. 4(A)) and the radiative 
corrections to the light--light scattering block, $M_{AR}$ (an example of
which is given in fig. 4(B)); $M_2$ denotes the remaining (non-annihilation
type) second order amplitudes.

We have seen that the contribution coming from the first order binding 
diagram, represents more than $90 \%$ of the first order radiative
corrections; hence it seems reasonable to look at second order corrections 
obtained by graphs containing some additional binding photon. 

In a recent paper \cite{io}, we have examined the contribution to the decay
width coming from the interference between the subtracted annihilation binding
graph represented in fig. 4(A) and the zero order diagram. In the same paper
we have also considered the square of the first order annihilation amplitude
${\cal M}_A$, contributing at $O(\alpha^2)$ to the decay rate, and we have
verified the existence of a logarithmically enhanced contribution arising from 
the radiative correction to the light-light scattering block depicted in 
fig. 4(B).     

Let's recall briefly how one can evaluate these different contributions
(for more details look at \cite{io}).

The matrix element of the annihilation binding diagram can be written as
(see fig. 4(A)):
\begin{equation}
{\cal M}_{AB}^{(m,\lambda)} = -\frac{i}{4 m^2} \, T_{\rho}^{(m)}\, 
G^{(\varepsilon)\rho} \, .
\label{matannbind}
\end{equation}

In the previous formula the tetravector $G^{(\varepsilon)\rho}$ describes 
the transition of the heavy photon to the three real ones and we have denoted 
with $\varepsilon \equiv \left(\varepsilon_1, \varepsilon_2, \varepsilon_3\right)$ 
the set of the three polarizations of these photons 
($\varepsilon_i = \pm 1$). 
The symbol $T_{\rho}^{(m)}$ represents the $O(\alpha)$ correction to
the annihilation current 4-vector of the positronium in the polarization state
$\varepsilon_m$.

One can easily see that $T_{\rho}^{(m)}$ contains a double integral, over 
the variables $p$ and $k$, a trace of $\gamma$ matrices, including also the
orthopositronium wave function $\Psi^{(m)} (p)$, and one photon propagator
that we can write as 
$ - \frac{i \ \Delta_{\mu \nu} (k - p)}{\left(k - p\right)^2}$.  

The $\Delta_{\mu\nu}$ tensor depends on the gauge we use. The choice
of the gauge is subtle when dealing with  bound state problems. It has been 
discussed elsewhere (see for instance the last paper of \cite{Oalfab}) 
that the Coulomb gauge is the most natural for calculations in positronium.
However, covariant gauges are simpler for 
computing radiative corrections, and, among them, the Fried--Yennie (FY) gauge
\cite{FYgauge} is the most convenient, due to its good infrared behaviour. 
We have computed $T^{(m)}_\rho$ both in the FY gauge and in the Coulomb 
gauge. As expected, the result is the same in both cases, and no gauge
correction term must be added when using the FY gauge.

Let's report the basic steps of the calculation in the FY gauge, for the
analogous computation in the Coulomb gauge we refer the interested reader to 
\cite{io}.
  
In the FY gauge we have $\Delta_{\mu \nu} = g_{\mu \nu} + 2 \frac{k_{\mu} 
k_{\nu}}{k^2}$. To perform the computation in this gauge, we have splitted the
trace entering $T_{\rho}^{(m)}$ into two pieces, one remaining non-singular at
$k=0$ and another one containing the contribution of the coulombic photon.
Formally we have used the following equality:
\begin{equation}
\Tr_{\mu \nu \rho} (k) = \Tr_{\mu \nu \rho} (0) + \left( \Tr_{\mu \nu
    \rho} (k) - \Tr_{\mu \nu \rho} (0) \right)\, . 
\label{splitFY}
\end{equation}

The first term gives a contribution to the matrix element proportional to the 
$O(\alpha)$ annihilation amplitude $\frac{\alpha}{\pi} {\cal M}_A$
\begin{equation}
{\cal M}_{A B, 1} = \frac{\alpha}{\pi} {\cal M}_A \left(1 - 3 \frac{\alpha}{\pi} \right)\, . 
\label{firstcontFY}
\end{equation}

The second term of eq.~(\ref{splitFY}) is infrared finite and we can safely 
put $p=0$ in the loop integral, introducing an error of order $O(\alpha^2)$. 
The ultraviolet divergence introduced by this second term can be regulated 
either with dimensional regularization or with the use of a cut-off. 

One must also add the contribution of the ``annihilation vertex''
counterterm, that cancels the divergence coming from the term 
$\left( \Tr_{\mu \nu \rho} (k) - \Tr_{\mu \nu \rho} (0) \right)$ as 
explicitly proved in \cite{io}. 
The sum of the contributions of the second term in eq.~(\ref{splitFY}) and of 
the counterterm gives:
${\cal M}_{A B, 2} = + \frac{\alpha}{\pi} \frac{\alpha}{\pi} {\cal M}_A$.

Hence we have found the following expression for the unsubtracted annihilation
binding diagram in the FY gauge:
\begin{equation}
{\cal M}_{A B} = \left(1 - 2 \frac{\alpha}{\pi} \right) \frac{\alpha}{\pi} 
{\cal M}_A \, .
\label{sumMABFY}
\end{equation} 

We must subtract the lowest order contribution $\frac{\alpha}{\pi} {\cal M}_A$
 corresponding to the exchange of a coulombic binding photon (as explained in 
the case of the $O(\alpha)$ binding correction) and consider the interference 
of this subtracted annihilation binding diagram with the zero order graph. 
In this way we have got the following $O(\alpha^2)$ contribution to the 
orthopositronium decay width:
\begin{equation}
\Gamma_{A B}^{'} = - 2 \frac{\alpha}{\pi} \Gamma_A = 1.6281
\left(\frac{\alpha}{\pi}\right)^2 \Gamma_0 \, , 
\label{bindanndecw}
\end{equation}
where in the last equation we have used the numerically improved value of the 
lowest order annihilation width, that can be found in the first paper of
\cite{vacpolar}: $\Gamma_A = -0.81406 \left(\alpha/\pi\right) \Gamma_0$\, .
This result is in very good agreement with the estimate, based on
factorization arguments, of this correction, that was made by Karshenb\u{o}im 
in \cite{Karsh2}.
 
In \cite{io} we have computed also the contribution to the decay width coming
from the square of the order $O(\alpha)$ annihilation amplitude 
(see fig. 3(A)). This square contains the integral over the phase space of the
final photons of the product of two tensors 
$G_{\rho}^{\varepsilon_1, \varepsilon_2, \varepsilon_3}$.   

To evaluate this quantity we have used the following relations (that can be
found in \cite{BFK} and in \cite{CDTP})
\begin{equation}
\int \frac{d^3 k_1 d^3 k_2 d^3 k_3}{\omega_1 \ \omega_2 \ \omega_3} 
\delta^{(4)} (P- k_1- k_2- k_3) = 8 \ \pi^2 \ m^2 \int d \nu_1\, d \nu_2\,  
d \nu_3 \delta (2 - \nu_1 - \nu_2 - \nu_3)
\label {relinteg}
\end{equation}
and
\begin{equation}
- \sum_{\varepsilon} G_{\rho}^{(\varepsilon)} G^{(\varepsilon) \rho} = \, 2^6 \
\alpha^4\ \left[R (234) + R (324) + R (423) \right]\, ,
\label{relGG}
\end{equation}
where $\nu_i = \frac{\omega_i}{m} = \frac{\left|\vec{k}_i\right|}{m}$ and 
$R$ are complicated functions of $\nu_i$ computed in \cite{CDTP} and reported 
in \cite{io}. 

Using eq.~(\ref{relinteg}) and eq.~(\ref{relGG}) we reduced the evaluation 
of the square of the annihilation amplitude to the calculation of a 
numerical integral. We have got the following numerical result:
\begin{equation}
\Gamma_{A^2} = (0  .17021 \pm 0.00010) \left( \frac{\alpha}{\pi} \right)^2 
\Gamma_0 \, .
\label{squarann}
\end{equation}

Let's notice that this result, like the annihilation-binding contribution 
of eq.~(\ref{bindanndecw}) (and differently from the lowest order 
annihilation contribution of $O(\alpha)$) has the right sign to reduce the 
eventual discrepancy between theory and experiment. 
Nevertheless, the absolute value of these corrections to the decay width 
is quite small and they are manifestly far from solving this discrepancy.
If the ``orthopositronium problem'' has to be solved by this kind of
perturbation theory, larger contributions must be searched in other classes 
of diagrams.

Finally, let' s remember that in \cite{io} we have also considered the
radiative correction to the light-light scattering block given by the graph
of fig. 4(B). It generates a logarithmically enhanced term that gives a
contribution to the decay width proportional to $\alpha^2 \ln (\alpha) \Gamma_0$.

To recover this result one can consider that there are two regions of the loop
momenta space giving the main contribution: one corresponding to $l, k \sim m$
and another one with fermion momenta almost on mass-shell, that is with $l, k 
\sim \alpha m$. The logarithmic term is produced in this second region.
In the analysis of this region one can put almost everywhere in the integral 
k and l to zero.
The integral over $l$ would be ultraviolet divergent, but, according to the
considerations we have just made, we can introduce the ultraviolet cutoff $m$,
and get the correct infrared logarithmic term. We have found a contribution to
the decay width that can be written as
\begin{equation}
\Delta \Gamma_{AR} = - \alpha^2 \ln \frac{1}{\alpha} \Gamma_0 \, ,
\label{radlog}
\end{equation}
in agreement with the results of \cite{radscattbloc}, where the all set of
these radiative corrections has been computed, and with the ones of
\cite{log1} and of the first paper of \cite{Oalfab}.  
                 
\vspace{0.7 cm}

{\sc{\bf Acknowledgments}}

I would like to thank all the other authors of the paper \cite{io}, from which
I have taken many of the new results reported here and in particular
V. Laliena. I am really grateful to him for many discussions that have been 
essential also for the preparation of this talk.
I am also indebted to P. Labelle for the very useful information and 
suggestions that he gave me.
It's a pleasure for me to thank all the organizers of the conference 
``Hadronic Atoms and Positronium in the Standard Model'', and in particular 
A. Rusetsky, for the
kind hospitality and for the unique human and scientific opportunity they
provided us.
Finally I would like to thank I.B. Khriplovich,S.G. Karshenb\u{o}im 
and A.S. Yelkhovskii for very useful discussions.
\vspace*{.6cm}

\begin{figure}[p]
\epsfxsize=10cm
\epsfysize=2cm
\epsffile{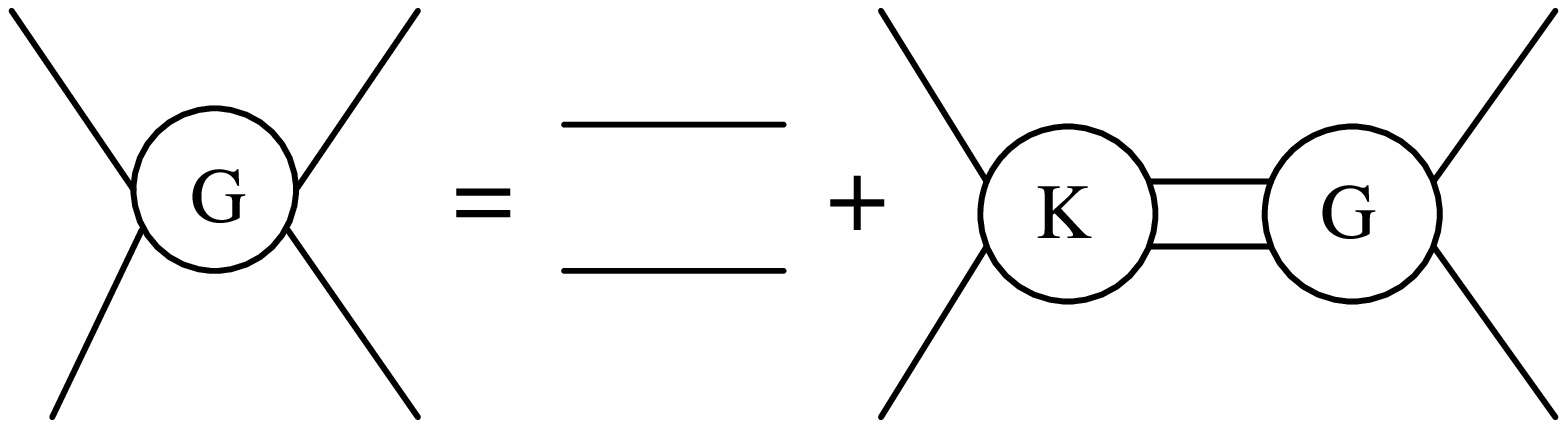}
\caption{Graphical representation of the Bethe-Salpeter equation. See the text
for the meaning of the symbols.}
\end{figure}

\begin{figure}[p]
\epsfxsize=9cm
\epsfysize=4cm
\epsffile{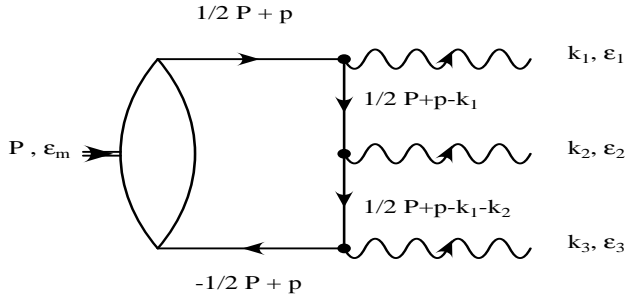}
\caption{Graph determining the orthopositronium decay width at lowest order.}
\end{figure}

\begin{figure}[p]
\epsfxsize=14cm
\epsfysize=4cm
\epsffile{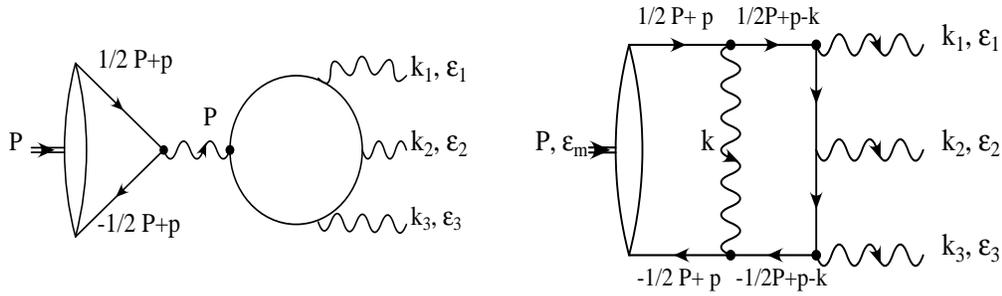}
\caption{Two different corrections of $O(\alpha)$:the annihilation graph
  and the binding diagram.}
\end{figure}

\begin{figure}[p]
\epsfxsize=14cm
\epsfysize=5cm
\epsffile{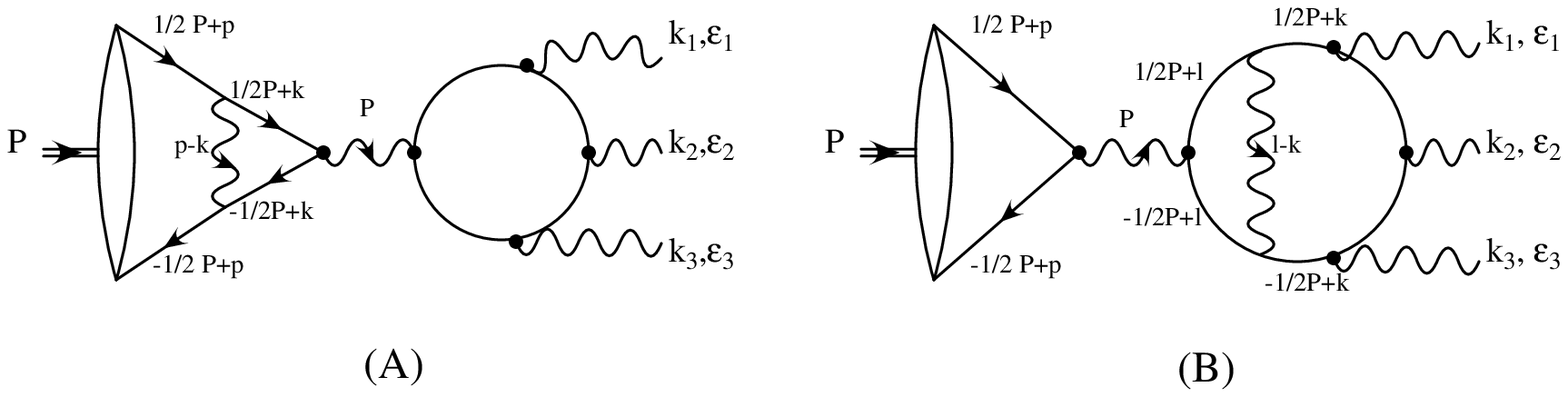}
\caption{Two different kinds of corrections to the annihilation graph.
(A) is the vertex correction, (B) represents 
the insertion of a photon into the light--light scattering block, which 
generates a logarithmically enhanced contribution. The direction of the 
fermions in the final loops is clockwise for all the two diagrams.}
\end{figure}



\begin{thebibliography}\

\bibitem{mic1}C.I. Westbrook, D.W. Gidley, R.S. Conti and A. Rich, {\it 
Phys. Rev. Lett.} {\bf 58} (1987) 1322; (Erratum {\it ibid} p. 2153);
 {\it Phys. Rev.} {\bf A40} (1989) 5489. 

\bibitem{mic2}J.S. Nico, D.W. Gidley, A. Rich and P.W. Zitzewitz, 
{\it Phys. Rev. Lett.} {\bf 65} (1990) 1344.

\bibitem{giap}S. Asai, S. Orito and N. Shinohara, {\it Phys. Lett.} {\bf B357}
(1995) 475.

\bibitem{mainz}P. Asbach, G. Hilkert, E. Klempt and G. Werth, {\it Nuovo 
Cimento} {\bf 97A} (1987) 419.

\bibitem{Ozero}A. Ore and J.L. Powell, {\it Phys. Rev.} {\bf 75} (1949) 1696.

\bibitem{Oalfa} W.E. Caswell, G.P. Lepage and J. Sapirstein, {\it
    Phys. Rev. Lett.} {\bf 38} (1977) 488. 

\bibitem {Oalfab} W.E. Caswell, G.P. Lepage, {\it Phys. Rev.} {\bf A20}
  (1979) 36;\\
G.S. Adkins, {\it Ann. Phys. (N.Y.)} {\bf 146} (1983) 78. 

\bibitem{OalfaAd} G.S. Adkins, A.A. Salahuddin and K.E. Schalm, {\it
    Phys. Rev.} {\bf A45} (1992) 7774. 

\bibitem{Adsolo} G.S. Adkins, {\it Phys. Rev. Lett.} {\bf 76} (1996) 4903.

\bibitem{log1} I.B. Khriplovich and A.S. Yelkhovskii, {\it Phys. Lett.} {\bf
    B246} (1990) 520.

\bibitem{log2} S.G. Karshenb\u{o}im, {\it Zh. Eksp. Teor. Fiz.} {\bf 103} 
(1993) 1105 [{\it JETP} {\bf 76} (1993) 541];\\ 
P. Labelle, G.P. Lepage and U. Magnea, {\it Phys. Rev. Lett.} {72} (1994) 
2006. 

\bibitem{io} V. Antonelli, V. Ivanchenko, E. Kuraev and V. Laliena, 
hep-ph/9706523, accepted for publication will appear on {\it
  Eur. Phys. Jour.} {\bf C} cod. DOI 10.1007/s100529800864 \, . 

\bibitem{bodyenrev} For a review on this subject see, for instance:\\
G.T. Bodwin and D.R. Yennie, {\it Physics Reports} (Section C of Phys. Lett.) 
{\bf 43 N.6 } (1978) 267.

\bibitem{Schwinger} J. Schwinger, {\it Proc. Nat. Acad. Sci.} {\bf US 37} 
(1951) 452.

\bibitem{Betesalp} E.E. Salpeter and H.A. Bethe, {\it Phys. Rev.} {\bf  84}
(1951) 1232.

\bibitem{Oalfaann} The most accurate determination of the $O(\alpha)$
  annihilation contribution can be found in \cite{OalfaAd}. This 
  contribution had been already studied, with increasing values of the
  magnitude and decreasing quoted uncertainty, in the the following papers:\\
M.A. Stroscio and J.M. Holt, {\it Phys. Rev.} {\bf A10} (1974) 749;\\
P. Pascual and E. de Rafael, {\it Lett. Nuovo Cimento} {\bf  IV} (1970) 1144;\\
W.E. Caswell, G.P. Lepage and J. Sapirstein, {\it Phys. Rev. Lett.} {\bf 38}
(1977) 488;\\
G.S. Adkins, {\it Ann. Phys. (N.Y.)} {\bf 146} (1983) 78. 

\bibitem{selfenvert} G.S. Adkins, A.A. Salahuddin and K.E. Schalm, {\it
    Phys. Rev.} {\bf A45} (1992) 3333.

\bibitem{sumsquares} A.P. Burichenko, {\it Yad. Fiz. } {\bf  56} (1993) 123
  [{\it Phys. At. Nucl.} {\bf 56} (5) (1993) 640]. 

\bibitem{radscattbloc} G.S. Adkins and M. Lymberopoulos, {\it Phys. Rev} 
{\bf A51} (1995) 2908.

\bibitem{vacpolar} G.S. Adkins and Y. Shiferaw, {\it Phys. Rev.} {\bf A52}
  (1995) 2442;\\
A.P. Burichenko and D. Yu. Ivanov, {\it Yad. Fiz. } {\bf 58} (1995) 898 [{\it
  Phys. At. Nucl.} {\bf 58} (1995) 832].

\bibitem{5fot} G.S. Adkins and F.R. Brown, {\it Phys. Rev.} {\bf A28} (1983)
1164;\\
G.P. Lepage, P.B. Mackenzie, K.H. Streng and P.M. Zerwas, {\it Phys. Rev.}
  {\bf  A28} (1983) 3090.

\bibitem{2ordrel} I.B. Khriplovich and A.I. Milstein, hep-ph/9607374 and BINP 
96-49;\\
I.B. Khriplovich and A.I. Milstein, {\it Zh. Eksp. Teor. Fiz.} {\bf 106} 
(1994) 689. [{\it Sov. Phys. JETP} {\bf 79} (1994) 379.]
 
\bibitem{Faustovetal} R.N. Faustov, A.P. Martynenko and V.A. Saleev, {\it 
Phys. Rev.} {\bf A51} (1995) 4520.

\bibitem{FYgauge} H.M. Fried and D.R. Yennie, {\it Phys. Rev} {\bf 112} (1958)
1391.

\bibitem{Karsh2} S.G. Karshenb\u{o}im, {\it Yad. Fiz.} {\bf  56} (1993) 155
  [{\it Phys. At. Nucl.} {\bf 56} (12) (1993) 1710].  

\bibitem{BFK} V.N. Baier, V.S. Fadin and E.A. Kuraev, {\it
    Sov. J. Nucl. Phys.} {\bf 31} (1980) 364.

\bibitem{CDTP} V. Costantini, B. de Tollis and G. Pistoni, 
{\it Nuovo Cimento} {\bf 2A} (1971) 733.


\end{thebibliography}
\end{document}